# O/IR Polarimetry for the 2010 Decade (CGT): Science at the Edge, Sharp Tools for All

A Science White Paper for the:
**Galaxies Across Cosmic Time (GCT) Science Frontiers Panel** of the
Astro2010 Decadal Survey Committee

Lead Authors:

Dean C. Hines
Space Science Institute
New Mexico Office
405 Alamos Rd.
Corrales, NM 87048
(505) 239 – 6762 (ph)
hines@spacescience.edu

Christopher C. Packham
Dept. of Astronomy
Univ. of Florida
211 Bryant Space Ctr
Box 112055
Gainesville, FL 32611
(352) 392-2052 x259 (ph)
packham@astro.ufl.edu

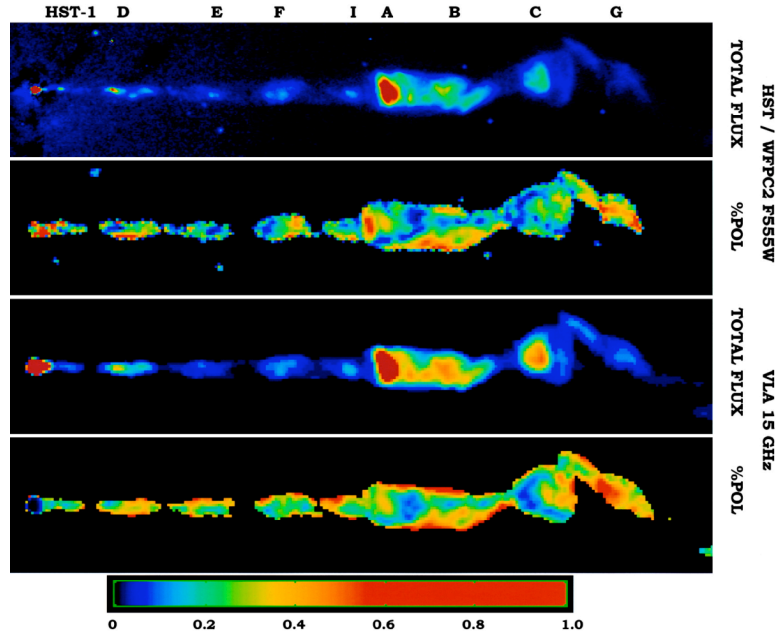

Total intensity and polarization of the M87 jet in the optical (*HST* F555W; *top two panels*) and radio (VLA 14.5 GHz; *bottom two panels*). After Perlman et al. (1999).

*Contributors and Signatories*

| | |
|---|---|
| Andy Adamson | UKIRT, JAC, Hilo |
| B-G Andersson | NASA Ames Res. Center |
| Robert Antonucci | University of California Santa Barbara, |
| David Axon | Rochester Institute of Technology |
| James De Buizer | SOFIA, NASA Ames |
| Alberto Cellino | Osservatorio Astronomico di Torino, Italy |
| Dan Clemens | Boston University |
| Jennifer L. Hoffman | University of Denver |
| Makoto Kishimoto | Max Planck Institut fuer Radioastronomie, Bonn |
| Terry Jay Jones | University of Minnesota |
| Alexander Lazarian | University of Wisconsin |
| Antonio Mario Magalhaes | University of Sao Paulo, Brazil |
| Joseph Masiero | University of Hawaii |
| Eric S. Perlman | *Florida Institute of Technology* |
| Marshall Perrin | UCLA |
| Claudia Vilega Rodrigues | Inst. Nac. De Pesquisas Espaciais, Brazil |
| Hiroko Shinnaga | CalTech |
| William Sparks | STScI |
| Doug Whittet | RPI |

## *Overview and Context: Polarimetry as a Cross-Cutting Enterprise*

Photometry, spectroscopy, and polarimetry together comprise the basic toolbox astronomers use to probe the nature of the universe. Polarimetry yields unique and powerful insight into a vast array of complex astrophysical phenomena, and has revealed the elusive magnetic field in the Milky Way and external galaxies, enabled mapping of features on unresolved stars, and was essential for establishing the Unified Model of Active Galactic Nuclei (in which all AGN have a fundamental structure consisting of a supermassive black hole and an accretion disk surrounded by a torus of dusty gas). Polarimetry is practiced across the full range of accessible wavelengths, from long wavelength radio through gamma rays, to provide windows into phenomena not open to photometry, spectroscopy, or their time-resolved variants. At some wavelengths, the U.S. leads the world in polarimetric capabilities and investigations, including ground-based radio with the VLA and VLBA. At other wavelengths, the U.S. is currently competitive: in submm the CSO and the JCMT have historically pursued similar science problems.

In ground-based O/IR, the situation is considerably worse, with no polarimeters available on Gemini or any NOAO-accessed 4 m (class) telescope. Over the past decade and more, Canadian and European astronomers have enjoyed unique access to state-of-the-art polarimeters and have used this access to vault far past the U.S. in many science areas.

| Telescope | Aperture | Instrument | Waveband | Polar. Mode | U.S. Access ? |
|---|---|---|---|---|---|
| IRSF (SAAO) | 1.4m | SIRPOL | NIR | Imaging | No |
| Perkins (Lowell) | 1.8m | Mimir, PRISM | NIR, Optical | Imaging | Private |
| HST | 2.4m | WFPC2, ACS, NICMOS | Optical, Optical, NIR | Imaging | Yes |
| Nordic Optical | 2.5m | TURPOL | Optical | Photopol | No |
| MMT | 6.5m | MMTPOL | NIR | Imaging | Private |
| LBT | 2x8.4m | PEPSI | Optical | Spectropol | Private |
| Gemini | 8m | Michelle | MIR | Imaging | Yes |
| Keck | 10m | LRIS | Optical | Spectropol | Private |
| GTC | 10m | CanariCam | MIR | Imaging | No |

In space, NICMOS, ACS and WFPC2[1] on *HST* have permitted imaging polarimetry at modest precision, and may represent the most general purpose UV/O/IR access for U.S. astronomers. Neither the *Spitzer Space Telescope* nor *JWST* have any polarimetric capabilities.

The dwindling U.S. access to this crucial third leg of the light analysis tripod has also become self-fulfilling, as students receive little exposure to polarimetric techniques and scientific advances as the number of practitioners able to teach students declines.

Nevertheless, polarimetric studies in the UV, optical, and IR have already revealed a great deal about star and planet formation processes, stars and their evolution, the structure of the Milky Way, and the nature and origin of galaxies and active galactic nuclei (AGNs) – details that cannot be obtained using pure photometric or spectroscopic methods. As we move into the next decade, we anticipate exciting advancement in areas such as the detailed physics of dusty circumnuclear tori, accretion disks and outflows/jets that comprise AGNs, as well as the intimate connections and interactions between AGNs and the galaxies that host them. The promise evident in the new, niche polarimetric instruments and the surveys they will perform will drive cutting-edge science in the upcoming decade. Yet teasing out answers to many key questions requires open community access to general purpose, precision polarimeters on large telescopes, as well as opportunities for student training.

---
[1] NICMOS and much of ACS are currently off-line until Servicing Mission 4 (SM4); WFPC-3 will replace WFPC-2 but will have no polarimetric capability.

## *Example Polarimetry Science Areas for the Next Decade*

Within the broad area covered by Galaxies Across Cosmic Time, polarimetric studies for the upcoming decade will focus on addressing key, and so far largely unsolved, problems in extragalactic astronomy, particularly with respect to AGNs: (1) How have AGNs evolved Across Cosmic Time, how do they influence their host galaxies, and do they dominate the re-ionization of the universe? (2) How do astrophysical jets form, propagate energy and matter, and influence the galactic and intergalactic environment? (3) What is the three-dimensional structure of central engines, including accretion disks and surrounding dusty tori? To tackle such questions, polarimetry provides unique capabilities that are not enabled by other techniques.

### The SMBH-Galaxy Connection Across Cosmic Time.

Over the last decade or so, it has become apparent that most galaxies contain supermassive black holes (SMBHs: e.g., Magorrian et al. 1998; Richstone et al. 1998), and that the SMBH mass ($M_{SMBH}$) is correlated with the larger scale mass of the galaxy ($M_{gal}$: e.g., see review by Kormendy 2004; all contributions in Ho 2004). What causes this intimate relationship, and if all galaxies have SMBHs, then why don't they all show AGN activity? One possible clue comes from investigations of ultraluminous infrared galaxies (ULIRGs), which appear to have a strong (evolutionary) link with AGNs (see review by Sanders 2004, and references therein). ULIRGs often show evidence of galaxy-galaxy interactions, with attendant star formation, and sometimes direct signs of AGN activity. Therefore, the interaction that triggers massive star formation may also feed the SMBH and energize the AGN. Is this the mechanism that drives the $M_{SMBH}$ *vs.* $M_{gal}$ correlation, or is there a different mechanism at work in the early universe? Polarimetry has played a pivotal role addressing these questions for low redshift objects by revealing the presence and nature of often directly obscured AGNs, even at very low luminosities. But we must move to higher redshifts where the real action is taking place, and this requires polarimetry instruments on big telescopes!

The results of such studies will also answer one of the key questions in early cosmology by identifying the dominant source of photons at the epoch of re-ionization: did stars or AGN re-ionize the early universe? We know that the basic structures of AGNs (accretion disks surrounded by dusty tori) are in place by z ~ 6 (~800 Myrs after the Big Bang: Hines et al. 2006; Jiang et al. 2006), and therefore we expect a substantial fraction of AGNs to be obscured from direct view. Polarimetry will not only identify such buried AGNs, but will also distinguish structures that are self-luminous from those illuminated by embedded light sources such as the central AGN engine. This observation can only be accomplished unambiguously with polarimetry, as was illustrated dramatically when polarimetry showed conclusively that elongated structures aligned with radio jets in some radio galaxies were dust scattered light from hidden QSOs as opposed to induced star formation as originally thought (Cimatti & di Serego Alighieri 1995). Here too, precision instruments with polarimetric capabilities on large telescopes will be crucial.

## AGN Physics: Accretion Disks, Dusty Tori and Jets.

*Accretion Disks* — Modeling the energetically-dominant UV/optical/near-IR continuum in AGNs has been stymied for decades, because the spectrum is very heavily contaminated by emission from broad lines, Balmer continuum and, longward of 1μm (rest frame), thermal emission from warm dust. No definite central-engine spectral features have been found in almost 50 years of taking total-flux spectra! But because of small-scale (interior to the Broad Line Region) electron scattering of the central engine light in otherwise unpolarized quasars, the central engine spectrum can be isolated using spectropolarimetry (Kishimoto et al. 2003; 2008). These polarized flux spectra are actually much more intelligible physically than the total flux spectra. The few results obtained so far provide an exciting proof of concept, which should now be exploited fully by studying accretion disk spectral energy distributions as function of $M_{SMBH}$, $L/L_{edd}$, radio jet properties, and host galaxy properties, all as a function of redshift. The biggest challenge for these studies will be access to high precision spectropolarimeters on large telescopes, and control of instrumental polarization to better than 0.1% in the JHK bands.

*Dusty Tori* — Infrared polarimetry holds the promise of investigating the geometrically and optically thick torus of gas and dust that surrounds the central engine in AGNs. The torus intercepts a significant fraction of the radiation from the central engine and the dust re-emits this energy in the MIR (3-30μm). The torus is compact (< few parsecs, e.g., Mason et al. 2006; Packham et al. 2005) and unresolved by even 8m telescopes, making it nearly impossible to *spatially* separate the torus emission from stellar and other dust emission from the host galaxy. Fortunately, NIR (1.2–2.2$\mu$m) polarimetry probes light scattered by dust near the torus, enabling structures on these size scales to be isolated, while mid-IR (3.8–5 $\mu$m) polarimetry probes polarized thermal emission from dust directly associated with the torus. In this case, dust grains in the torus are aligned by radiative torques or some other alignment mechanism (see review by Lazarian 2007), which polarizes this thermal emission. Therefore, MIR polarimetry, in particular, offers a powerful and largely untapped way to isolate the torus emission (e.g., Packham et al. 2007).

*Jets* — The origin and physics of relativistic jets associated with AGNs is also among the most compelling current problems in astrophysics. Magnetic fields (B-fields) play a prominent role in the physical processes that occur in AGN jets, and the leading model for jet production, acceleration, and collimation involves poloidal B-fields that are wound up by the differential rotation of a rotating disk or ergosphere surrounding a central SMBH (e.g., McKinney & Narayan 2007a,b; Meier et al. 2001). Because the primary emission mechanism from radio through optical wavelengths is synchrotron radiation, linear polarimetry probes directly the configuration of the B-fields in the emission regions. Polarization observations at all wavebands are important, but especially so in the mid- and far-IR part of the spectrum, where the synchrotron emission arises from the zone where the jet is collimated and accelerated to relativistic speeds. Advances in IR polarization capabilities will thus probe the B-field geometry in this zone, thereby testing directly the theories of jet formation.

Beyond the acceleration region — which may extend over hundreds or thousands of gravitational radii from the SMBH (Vlahakis & Konigl 2004; Marscher et al. 2008) — the jet may become turbulent or subject to velocity shear. The geometry and degree of order of the B-field (as measured with polarimetry) are therefore key indicators of the physical conditions in a jet. These jets also terminate within the interstellar or the intergalactic medium (ISM & IGM), where they deposit energy and material that can have profound influence on the ISM and IGM. In some cases, filaments and bubbles are observed at termination zones, but their exact nature is very uncertain (see, e.g., M87: Forman et al. 2007). Optical and infrared polarimetry of these features distinguishes their nature as, for example, synchrotron filaments associated with shocks and turbulent magnetic fields, or scattered (polarized) light from density enhanced material illuminated by the central engine. Results from such studies will have a profound impact on our understanding of the feedback and influence of AGNs on their environments, with implications ranging from the suppression cooling flows in galaxy clusters (e.g., Sijacki et al. 2007), to the origins and amplification of cosmic magnetic fields (see, e.g., Rees 2005, 2006).

*Exceptional Discovery Potential Area:* Polarimetric Tomography of AGNs

Mapping the 3D structure of AGN ("AGN Polarimetric Tomography") is arguably one of the the most exciting prospects for UV/Optical/IR (spectro-) polarimetry of AGNs in the coming decade. Typically, astrophysical tomography relies exclusively on Doppler emission line imaging of time-variable, usually rotating, systems, but polarimetry provides tomography for a much broader range of astrophysical situations. In particular, light polarized by scattering can provide views of a given structure from multiple (i.e., different) vantage points (i.e., as viewed from each scattering cloud), and this information enables construction of a detailed 3D picture of both the scattering clouds and the illuminating source(s). This extraordinary capability exploited for objects over a range of redshfts, has the exciting potential to show us the *three dimensional* evolution of AGNs across cosmic time. The potential of this technique for galaxy research was anticipated early on (e.g., M82: Schmidt et al. 1976), but advancing technology in polarimeter design and access to larger telescopes is finally enabling application to a broad range of problems, and the method is attracting increased attention (Miller et al. 1991; Smith et al. 2005).

As an exciting example, pioneering modeling of *HST* UV polarimetric observations of the AGN, NGC1068 by Kishimoto (1999) produced a stunning 3D image of the central region, as well as locating the astrometric position of the AGN with unprecedented precision. Buttressed by previous spectropolarimetry (Miller et al. 1991), Kishimoto assumed that the degree of polarization is wholly a function of the angle between (a) our line of sight to the scattering agent and (b) the line of sight of those scattering agents to the AGN. Thus the location of the scattering cloud relative to the AGN can be determined. In this way, a 3D view of the nuclear regions of NGC1068 can be constructed, at least for those clouds that have pure scattering from the AGN. The result constitutes, at least to the authors' knowledge, the first polarimetricly determined 3D image of the central region of an AGN. The polarized emission (i.e., scattered light) is extended along a linear type feature, which is probably associated with the well-known radio jet (Figure 1, below).

Finally, we emphasize that the polarized spectrum can isolate the AGN from all other (unpolarized) emission regions. Therefore, for objects that are not spatially resolved, spectropolarimetry has the (hitherto untapped) potential to directly measure "reverberation events" within the accretion and emission line regions in the central engine without contamination from other emission sources (e.g., Goosman & Gaskell 2007; Goosman et al. 2008). Therefore, the prospect of reconstructing the 3D structure of the AGN even at the highest redshifts is within our grasp if we have the necessary, accessible instrumentation on large telescopes!

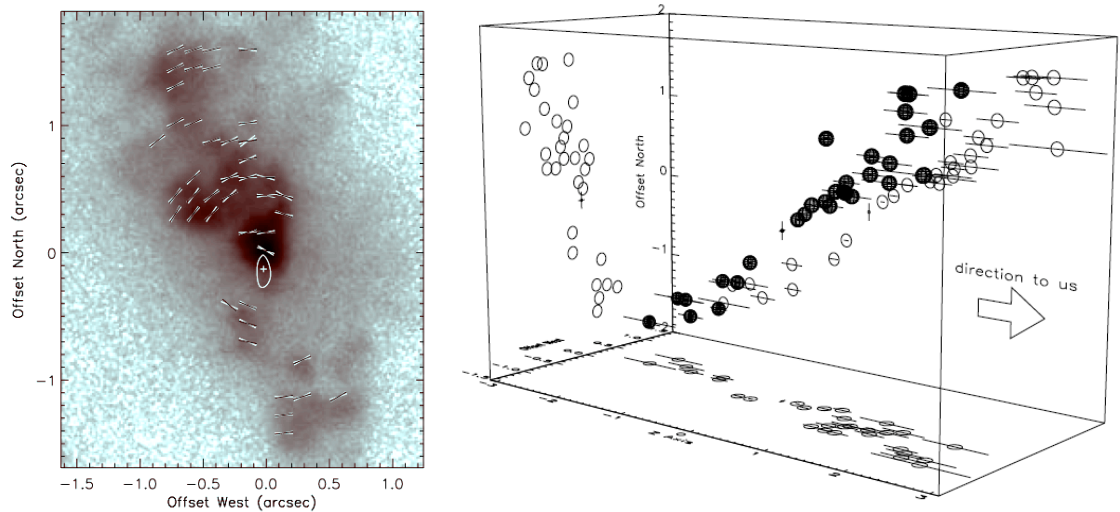

**Figure 1**: The UV polarization of NGC1068 with error bars (left), and the 3-dimensional scattering plot (right). All figures taken from Kishimoto (1999).

## *What is Needed to Meet These Science Goals within the Decade*

What key observations and instrumentation are needed to achieve the science goals within the next decade? Our evaluation of the upcoming opportunities, challenges, and technical readiness leads to the following recommendations:

*1) Build precision polarimetric capability into new (UV)/O/IR instruments for large telescopes and space missions. Design polarimetry in from the beginning, not as "add-ons".*

   a. New "niche" instruments, such as GPI, SPHERE/ZIMPOL, and HiCIAO on Subaru rightly exploit polarimetry to meet their exoplanet objectives, but general purpose instruments with polarimetric capability are lacking at virtually all large, open-access US telescopes - ground-based, airborne, and space-based, especially in the infrared.
   b. Key science questions cannot be answered unless US astronomers have access to precision (photon-noise limited) polarimetric capability. To retain precision capability, polarimetric capabilities must be a considered in the initial design of the instruments, not as a later "add-on".

*2) Develop polarimetric O/IR capabilities on intermediate-size telescopes to continue investigate nearby galaxies and AGNs, to promote student training in instrumentation and polarimetric observations.*

   a. To be able to compete scientifically, we must invest in the next generation of young astronomers who will use polarimetry as a powerful tool in their light analysis toolbox and who will understand polarimetric light analysis well enough to guide future instrument development.

b. Obtaining ground-based calibrating polarimetric observations are crucial to the calibration of space-based polarimeters. This has been increasingly important, especially for ACS and NICMOS aboard *HST*. Current limitations of existing standards (e.g., sparse, too bright, few good IR targets, etc.) would be solved using ground-based telescopes.

## *Final Thought*

The U.S. astronomical community has lost opportunities to advance key science areas as a result of down-selects of instrument capabilities or lack of will to commission polarimetric modes on instruments. The investment is minor (especially if the option for polarimetry is included in the baseline design), the expertise is available in the community, and the rewards are tangible. We are excited by the recent momentum favoring polarimetric studies and capabilities, and believe the upcoming decade will see the various polarimetric techniques together become a strong, necessary component of astronomers' light analysis toolbox.

## *Bibliography and References*